\documentclass[pre,eqsecnum,preprint,showpacs,floatfix]{revtex4}
\usepackage{amssymb,amsmath,bm,graphicx}
\font\bba=msbm10 scaled 1080
\font\bbb=msbm8 
\font\bbc=msbm6 
\newfam\bbfam
\textfont\bbfam=\bba
\scriptfont\bbfam=\bbb
\scriptscriptfont\bbfam=\bbc
\def\bb{\fam\bbfam\bba}
\def\R{{\bb R}}

\begin{document}
\title{Maxwell field $ \mathbf{E}$  in a two-dimensional polar fluid
 in the presence of an external field $\boldsymbol{\mathcal{E}} $ :  a Monte-Carlo study}
\author{Jean-Michel Caillol}
\email{Jean-Michel.Caillol@th.u-psud.fr}
\author{Dominique Levesque}
\email{Dominique.Levesque@th.u-psud.fr}
\author{Jean-Jacques Weis}
\email{Jean-Jacques.Weis@th.u-psud.fr}
\affiliation{Laboratoire de Physique Th\'eorique \\
CNRS (UMR 8627), B\^at. 210\\
Universit\'e de Paris-Sud \\
91405 Orsay Cedex, France}
\date{\today}
\begin{abstract}
We study a two-dimensional system of dipolar hard disks in the presence of a uniform 
external electric field  $\boldsymbol{\mathcal{E}} $ by Monte Carlo simulations
in a square with periodic boundary conditions. The study is performed in both
the fluid at high temperature and the phase of living polymers at low temperature.
In the considered geometry the macroscopic Maxwell field $ \mathbf{E}$ is computed
and found to be equal to 
the external field  $\boldsymbol{\mathcal{E}} $ in both phases. The dielectric  properties of the system
in the liquid phase as well as in the polymeric phase are investigated.
\end{abstract}
\maketitle
\newpage
\section{Introduction}
\label{intro}
We dedicate our contribution to the memory of George Stell with whom the Loup Verlet's Orsay group has shared
a long-lasting friendship.

This study is devoted to a Monte-Carlo (MC) simulation of a two-dimensional (2D) system made of identical dipolar hard disks (DHD) in the Euclidian plane $E_2$ 
in the presence of an uniform  external electrostatic field $\boldsymbol{\mathcal{E}}$.
The dipoles are assumed to be permanent and the configurational energy of
$N$ dipolar molecules in  $E_2$ reads as
\begin{eqnarray}
\label{I1}
H &= &
\frac{1}{2}\sum_{i \ne j}^{N} v_{HS}(r_{ij}) 
+ \frac{1}{2} \mu^2 \sum_{i \ne j}^{N}
 \frac{1}{r_{ij}^2}   \left[ \mathbf{s}_i \cdot
\mathbf{s}_j - \frac{2(\mathbf{s}_i \cdot \mathbf{r}_{ij} )
 (\mathbf{s}_j \cdot \mathbf{r}_{ij} )}{ r_{ij}^{2}}\right]  \nonumber \\
& - &\mu \sum_{i =1 }^{N} \mathbf{s}_i \cdot \boldsymbol{\mathcal{E}}
\nonumber  \\  \; .
\end{eqnarray}
In Eq.\ \eqref{I1},  $v_{HS}(r)$ is the hard disk potential  of diameter $\sigma$.
The second term is the  contribution from the 2D dipole-dipole interaction
where ${\bm \mu}_i = \mu \mathbf{s}_i$, $\mu$ permanent dipole
moment, $\mathbf{s}_i$  unit vector in the direction of the
dipole moment of  particle $i$,
$\mathbf{r}_{ij}= \mathbf{r}_{j} - \mathbf{r}_{i}$, the vector joining the centres of mass 
of the particles, and $ r_{ij}= |\mathbf{r}_{ij}|$.  Finally, the  last term denotes the interaction
energy of the dipoles with the external field $\boldsymbol{\mathcal{E}} $ which is assumed
to be uniform.
We note that the dipole-dipole interaction involved in Eq.~\eqref{I1} is derived from the solution of the 2D Laplace equation in the plane. 
This model was studied recently in the absence of $\boldsymbol{\mathcal{E}} $
 by means of MC simulations performed on a sphere and
in a square with periodic boundary conditions (PBC)~\cite{Caillol_Weis}. In both cases
the laws of electrostatics governing the interactions in the considered geometries  were scrupulously adopted
yielding identical phase diagrams.
At high temperature and moderate density an ordinary polar fluid
is observed characterized by a dielectric constant $\epsilon$.
In the low temperature, low density part of the phase diagram
a phase of living polymers of dipoles organized into closed rings has been observed. At higher density the 
structure of this phase is characterized by an entangled structure of chains and rings and 
the dielectric constant $\epsilon$
is no more well-defined. It was found that the critical dipole moment  $\mu_c^*$ at the
transition from fluid to polymeric phase increases slightly
with density.

In the presence of an applied field, in the low temperature phase the dipole moments are  expected to align in the direction of the 
field, the tendency getting more effective as the strength of $\boldsymbol{\mathcal{E}} $ increases,
first yielding roughly linear chains which ultimately will collapse into bundle-like structures at high density.
As it is impossible to consider uniform fields on the sphere (without
violating the laws of electrostatics, cf.~\cite{Caillol}) we present here only simulations performed
of DHD contained in a square  with  PBC.

\section{Electrostatics of $2D$ dipoles within periodical boundary conditions}
\label{Simu}
Let us first recall some elements  of electrostatics for
a square of side  $L$ along directions $Ox, Oy$
with PBC, which  will be referred to as space $\mathcal{C}_2$~\cite{Caillol_Weis,perram:81,morris:85}.
\subsection{The dipolar Green's function in $\mathcal{C}_2$}
\label{Green}
The electric field
at point  $\mathbf{r}_1=x_1 \mathbf{e}_x + y_1\mathbf{e}_y$ ($ (x_1,x_2)  \in [-L/2,L/2]$, $(\mathbf{e}_x ,  
\mathbf{e}_y)$ orthonormal basis of $\mathcal{C}_2$)
created by a point dipole $ {\bm \mu}_2$ located at point $\mathbf{r}_2=
x_2 \mathbf{e}_x + y_2\mathbf{e}_y$  of
$\mathcal{C}_2$ is given by $2 \pi \mathbf{G}_0 (\mathbf{r}_1,\mathbf{r}_2) \cdot {\bm \mu}_2$ where
$\mathbf{G}_0$ denotes the bare dipolar Green's function and the dot  a tensorial contraction (see \textit{e.g.}
Refs.~\cite{Caillol,Fulton_I,Fulton_II,Fulton_III}).

The Green's function $\mathbf{G}_0$ depends on the considered geometry and is given in $\mathcal{C}_2$ by
\begin{equation}
 \mathbf{G}_0(\mathbf{r}_1, \mathbf{r}_2) \equiv \mathbf{G}_0(\mathbf{r}_{12}) =
  \dfrac{1}{2 \pi}\dfrac{\partial}{\partial \mathbf{r}_{12}} 
\dfrac{\partial}{\partial \mathbf{r}_{12}}  \psi(\mathbf{r}_{12}) \; ,
\end{equation}
where  $\psi(\mathbf{r})$ is the periodic Ewald potential.
 The latter  satisfies Poisson's equation in  $\mathcal{C}_2$,
\textit{i.e.}
\begin{equation}
 \Delta \psi(\mathbf{r}) = -2 \pi [\delta_{\mathcal{C}_2}  ( \mathbf{r})     - \dfrac{1}{L^2} ] \; ,
\end{equation}
where 
\begin{align}\label{toto}
 \delta_{\mathcal{C}_2}  ( \mathbf{r})= &\sum_{\mathbf{n}} \delta^{(2)}(\mathbf{r} -L \mathbf{n}) \, , \nonumber \\
=& \dfrac{1}{L^2}\sum_{\mathbf{k}} \exp (i\mathbf{k} \cdot  \mathbf{r})  \, ,
\end{align}
is the periodical Dirac's comb. In  Eqs.~\eqref{toto} $\mathbf{k}= (2 \pi/L ) \mathbf{n}$ where $\mathbf{n}$ 
is a $2D$ vector with integer components $ (n_x,n_y)$ in the basis $(\mathbf{e}_x ,  
\mathbf{e}_y)$ . One notes that $\psi(\mathbf{r})$ identifies with the potential
created by a unit point charge immersed in a uniform neutralizing background of charge density $-1/L^2$.
 Expanding  $\psi(\mathbf{r})$ and 
$\mathbf{G}_0(\mathbf{r})$ in Fourier series
one finds
\begin{subequations}
 \begin{align}
 \psi(\mathbf{r}) =& \dfrac{2 \pi}{L^2} \sum_{\mathbf{k} \neq \mathbf{0} }\dfrac{\exp(i\mathbf{k} \cdot  \mathbf{r})}{\mathbf{k}^2} \, , \\
 \mathbf{G}_0(\mathbf{r}) =& - \dfrac{1}{L^2} \sum_{\mathbf{k} \neq \mathbf{0}} \widehat{\mathbf{k}} \widehat{\mathbf{k}}   \exp(i\mathbf{k} \cdot  \mathbf{r}) \; ,\label{G0}
 \end{align}
\end{subequations}
where  $\widehat{\mathbf{k}} =  \mathbf{k}/\Arrowvert  \mathbf{k}  \Arrowvert$.
Some comments on the previous developments are in order:
\begin{itemize}
 \item{(i)} 
  It follows from Eq.~\eqref{G0} that
\begin{equation} \label{IntG0} 
 \int_{\mathcal{C}_2}\; d\mathbf{r}_2 \, \mathbf{G}_0(\mathbf{r}_1, \mathbf{r}_2) =\mathbf{0} \; , 
\end{equation}
and therefore the electric field created by a uniform polarization $\mathbf{P}$ is zero (see Sec.~\eqref{Maxwell}).
\item{(ii)}  Clearly $\mathbf{G}_0(\mathbf{r})$ exhibits the same singularity for  $\mathbf{r} \to \mathbf{0}$ as
the Green's function of the infinite Euclidean plane $\R^2$, \textit{i.e.}\cite{Caillol,Fulton_I,Fulton_II,Fulton_III,Jackson} 
\begin{subequations}\label{decompo} 
 \begin{align} 
 \mathbf{G}_0(\mathbf{r}) & =   \mathbf{G}_{0}^{ \delta} (\mathbf{r}) - \, \frac{1}{2} \,\delta(\mathbf{r}) \,     
                                             \mathbf{U} \; ,  &  \label{decompo_a}      \\
  & \mathbf{G}_{0}^{ \delta} (\mathbf{r})= \begin{cases}
                                                 \mathbf{G}_{0} (\mathbf{r}) \; &,   \text{ for } r > \delta  \; ,  \\        
                                                  0                        \;  &,    \text{ for } r < \delta  \; , 
                                            \label{decompo_b}          \end{cases}
 \end{align}
\end{subequations}
where $\mathbf{U}=\mathbf{e}_x\mathbf{e}_x + \mathbf{e}_y \mathbf{e}_y $ is the unit dyadic tensor.
In Eqs.~\eqref{decompo} $\delta$ is an arbitrary small cutoff ultimately set to zero if point dipoles are to be considered.
It must be understood that any integral
involving $\mathbf{G}_0^{\delta} $ must be calculated with $\delta \neq 0$
and then taking the limit $\delta \to 0$. In the presence of hard cores, as in the DHD fluid, the point dipoles
can be replaced by a continuous charge distribution of symmetry axis $\mathbf{s}$ and charge density
$\propto \mathbf{s}\cdot \mathbf{r}$.
In that case $\delta$ can be  chosen to take any value $0<\delta \leq \sigma/2$.
\item{(iii)} 
The interaction energy of two dipoles of $\mathcal{C}_2$ is given by $- 2 \pi {\bm \mu}_1
 \cdot \mathbf{G}_0 (\mathbf{r}_1,\mathbf{r}_2) \cdot {\bm \mu}_2$.
As well-known, expression~\eqref{G0} cannot be handled in a straightforward way in MC simulations since it converges
very slowly. This Fourier series is then splitted into two series, one in direct space, the other
in Fourier space, both with good convergence properties. In that way  one obtains the dipolar Ewald potential
which is detailed in Ref.~\cite{Caillol_Weis} and which was used in the simulations reported in this paper.
\end{itemize}
\subsection{Linear distributions of dipoles in $\mathcal{C}_2$} 
We consider a line of length $L$, parallel to axis $Oy$ of a square with PBC,  which bears a continuous distribution of dipoles $d\bm{\lambda}=(\lambda_x\mathbf{e}_x+
\lambda_y \mathbf{e}_y)dy$ where the components $(\lambda_x,\lambda_y)$ are constants. Without loss of generality 
the equation of the line can be chosen as $x=0$. The electric field  $d\boldsymbol{\mathcal{E}} $ created by this infinitesimal dipole  
 at point $\mathbf{r}$ is given by
\begin{align}\label{dedip}
 d\boldsymbol{\mathcal{E}} &=  2 \pi \mathbf{G}_0(\mathbf{r}) \cdot d\bm{\lambda} \, , \nonumber \\
                                           &= - \frac{2 \pi} {L^2}  \sum_{\mathbf{k} \ne \mathbf{0}}\; \frac{\mathbf{k} \mathbf{k} }{k_x^2 + k_y^2} 
                                              \exp(i (k_x x + k_y y)) \cdot (\lambda_x \mathbf{e}_x+ \lambda_y \mathbf{e}_y)dy\, , \nonumber  \\
                                           &= - \frac{2 \pi} {L^2}  \sum_{\mathbf{k} \ne \mathbf{0}}\;  \frac{\mathbf{k} }{k_x^2 + k_y^2}    \exp(i (k_x x + k_y y))
(\lambda_x k_x +
\lambda_y k_y)dy \, .
\end{align}
The total field of the line is obtained by integration

\begin{align} 
\label{edip}
 \boldsymbol{\mathcal{E}}  & = 2 \pi \int_{-L/2}^{L/2}   d\boldsymbol{\mathcal{E}}  \, . \nonumber \\
                                           &= - \frac{2 \pi \lambda_x} {L^2}   \sum_{\mathbf{k} \ne \mathbf{0}}\;  \frac{ k_x \mathbf{k}}{k_x^2 + k_y^2}    \exp(i k_x x) L \delta_{k_y,0}\, , \nonumber  \\
                                           &= - \frac{2 \pi \lambda_x} {L} \sum_{k_x \ne 0} \exp(i k_x x )  \mathbf{e}_x \; , \nonumber \\
                                           &= -2 \pi \lambda_x \left[ \delta(x) - \frac{1}{L} \right]  \mathbf{e}_x  \; .
\end{align}

Eq.~\eqref{edip} shows that the field of a uniform distribution of dipoles aligned along the line $x=0$ ($\lambda_x=0$) 
vanishes exactly. This result should be compared with the expression of the field
created by an infinite linear chain of point dipoles, each with dipole moment $\mu \mathbf{e}_y$, all  aligned along the $Oy$ axis with spacing
$2a$ given in Ref.~\cite{Toor}. The latter is expressed as a series, the dominant term of which behaves
as $\propto (\mu/a^2) \exp(-\pi \vert x \vert/a) \cos(\pi y /a)$. With the correspondence $\mu = 2 \lambda_y a$ and in the limit of a continuous distribution, \textit{i.e.}
$a, \mu \to 0$ with $\lambda_x$ fixed, this dominant term vanishes in agreement with our result.
These results suggest that in the polymeric phase and  in the presence of a sufficiently large external field the chains of dipoles
aligned along the field do not contribute significantly to the Maxwell field. 
This point has been checked in our simulations and is discussed in Sec.~\eqref{MCdat}.

 The expression~\eqref{edip} of $ \mathcal{E}$ also shows that, apart from
the singularity  $-2 \pi \lambda_x \delta(x) \mathbf{e}_x$, a uniform electric field parallel to the axis $Ox$ of strength $2 \pi \lambda_x/L$  
can be generated by a uniform distribution 
of dipoles perpendicular to the line ($\lambda_y=0$). In passing we note that the delta singularity has no effect in actual simulations since the condition $x=0$,
of measure zero, is
never obtained. We also remark that $\lambda_x$ being fixed,  the field vanishes  in the limit $L \to \infty$  as it should be, since then the result of  the Euclidean plane $\R^2$ with free boundaries at infinity
should be recovered. The electrostatic potential obtained by integration, is found,
up to an arbitrary additional constant, to be
\begin{align}
 V(x) &= \begin{cases}
                 =  \pi \lambda_x    - \dfrac{2 \pi}{L}  \lambda_x x  \text{ , for } x> 0       \\
                 = -  \pi \lambda_x  - \dfrac{2 \pi}{L}  \lambda_x x  \text{ , for } x< 0  \, ,
         \end{cases}
\end{align}
yielding the expected discontinuity of the potential across a dipole layer\cite{Jackson}
\begin{equation}
 V(0+) - V(0-) = 2 \pi \lambda_x  \, .
\end{equation}
These results can be extended to the $3D$ case (with $2 \pi$ replaced by $4 \pi$) where one should consider instead a 
planar uniform layer
of dipoles aligned along the normal to the square  $x=0$ of the cube $\mathcal{C}_3$. 
We stress that it seems to be  the only way to generate a uniform
electrostatic  field within PBC cubic geometries.
\subsection{The Maxwell field in $\mathcal{C}_2$}
\label{Maxwell}
At thermal equilibrium the Maxwell field $\mathbf{E}(\mathbf{r})$ is the sum of the external field $\boldsymbol{\mathcal{E}}(\mathbf{r})$
and the field created by the dipoles $\mathbf{E}_d(\mathbf{r}) = <\widehat{\mathbf{E}}_d(\mathbf{r})>$. It follows from Sec.~\eqref{Green}
that, in a given configuration of  $N$ dipoles in the canonical ensemble, the microscopic field $ \widehat{\mathbf{E}}_d(\mathbf{r})$ is given by
\begin{equation}
  \widehat{\mathbf{E}}_d(\mathbf{r}) = 2 \pi \int_{\mathcal{C}_2} d\mathbf{r}^{'} \mathbf{G}_0(\mathbf{r}, \mathbf{r}^{'}) \cdot
\widehat{ \mathbf{P}}(\mathbf{r}^{'} ) \; , 
\end{equation}
where the microscopic polarization at point $\mathbf{r}$ reads
\begin{equation}
 \widehat{\mathbf{P}}(\mathbf{r}) = \sum_{i=1}^N \boldsymbol{\mu}_i \delta_{\mathcal{C}_2}(\mathbf{r} - \mathbf{r}_i) \; .
\end{equation}
At equilibrium we thus have 
\begin{equation}
  \mathbf{E}(\mathbf{r}) =      \boldsymbol{\mathcal{E}}(\mathbf{r}) +        2 \pi \int_{\mathcal{C}_2} d\mathbf{r}^{'} \mathbf{G}_0(\mathbf{r}, \mathbf{r}^{'}) \cdot
 \mathbf{P}(\mathbf{r}^{'} ) \; , 
\end{equation}
where  the macroscopic polarization $\mathbf{P}(\mathbf{r}) = <   \widehat{\mathbf{P}}(\mathbf{r}) >$.

For a fluid in a uniform external field $\boldsymbol{\mathcal{E}}$ the polarization $\mathbf{P}$ is uniform
and, for a sufficiently low field strength,   $2 \pi \mathbf{P} = (\epsilon -1)\mathbf{E} $, which defines the dielectric constant $\epsilon $.
As well known~\cite{Jackson} the relation between the Maxwell field and the external field depends on the geometry. 
For instance, in the 
plane $\R^2$ with free boundaries at infininity we have $ \mathbf{E}=  \boldsymbol{\mathcal{E}}/\epsilon$ while,
in $\mathcal{C}_2$ we have
\begin{align}
 \mathbf{E}_d &=  \left[ 2 \pi \int_{\mathcal{C}_2} d\mathbf{r}^{'} \mathbf{G}_0(\mathbf{r}, \mathbf{r}^{'})\right] \cdot \mathbf{P} \, ,\nonumber \\
                      &= \mathbf{0} \; ,
\end{align}
as a consequence of Eq.~\eqref{IntG0} from which it follows that $\mathbf{E} = \boldsymbol{\mathcal{E}}$ and
therefore $2 \pi \mathbf{P} = (\epsilon -1)\boldsymbol{\mathcal{E}}$.
\section{MC data analysis}
\label{MCdat}
For the present planar system the MC simulations were performed in a square
 of surface $A=L^2$ with PBC for $N=1024$ hard disks 
carrying a dipole moment, in reduced units, $\mu^*= \mu/\sqrt{kT\sigma^2}$
varying between $\mu^*=1.5$ and $2.75$.
The densities, in reduced units, were taken in the range $\rho^*=\rho
\sigma^2$ from 0.1 to 0.3.
The energy $E$ of the system in the presence of the external field ${\bf \cal E}$
is calculated, taking into account the PBC, by the Ewald summation technique
explicated in Refs \cite{Caillol_Weis,perram:81,morris:85}. 
The expression at point ${\bf r}$ of the microscoipic field $\widehat{{\bf E}}_d({\bf r})$ due to the $N$ dipoles ${\boldsymbol \mu}^*_i=\mu^* {\bf s}_i$ 
is thus given by
   \begin{eqnarray}  
 \widehat{{\bf E}}_d({\bf r})  &=& 2 \pi \sum_{i=1}^N \mathbf{G}_0 (\mathbf{r},\mathbf{r}_i) \cdot {\bm \mu}^*_i \nonumber \\
                           & \simeq & \sum_{i=1}^N \Big\{ {({\bf r - r}_i)} {\frac {2 \, \exp(-\alpha^2 |{\bf{r}-\bf{r}_i}|^2)}{|{\bf{r}-\bf{r}_i}|^2}}
                         ({\frac{1}{|{\bf{r}-\bf{r}_i}|^2}+\alpha^2}) ({\bf{r}-\bf{r}_i}).{\bm \mu}^*_i 
                          - {\frac {\exp(-\alpha^2 |{\bf{r}-\bf{r}_i}|^2)}   
                           {|{\bf{r}-\bf{r}_i}|^2}}{\bm \mu}^*_i \Big\} \nonumber \\ 
     & +& {\frac{ 2\, \pi}{A}} \sum_{{\bf k \neq 0}}{{\bf k}} 
    { \frac { \exp(-|{\bf k}|^2/(4\alpha^2))}{|{\bf k}|^2}}\sum_i 
   \exp(-i{\bf k}.{\bf r}_i)\exp(i{\bf k}.{\bf r}){{\bm \mu}^*_i}.{\bf k}  \, ,
        \label{eq4}  
    \end{eqnarray}
where we have reported in the second line of Eq.~\eqref{eq4} the Ewald
expression used in our simulations. The parameter $\alpha$ regulates the rate
of convergence of the sums in direct and Fourier space.

At equilibrium the Maxwell, or macroscopic field is defined as 
\begin{equation}\label{Maxwell_bis}
 \mathbf{E}= \boldsymbol{\mathcal{E}} + \left\langle \widehat{{\bf E}}_d \right\rangle \, ,
\end{equation}
where the brackets denote a canonical average, while the macroscopic polarization
is the canonical average of $\widehat{{\bf P}} ({\bf r})=  \mu^*  
 \sum_i {\bf s}_i \delta ({\bf r}-{\bf r}_i) $. In the region of the phase diagram
where it is a defined quantity  and in the limit of small fields, the dielectric constant
 $\epsilon$ of the system can be obtained, as discussed in Sec.~\eqref{Maxwell}, through  $
 2 \pi <\widehat{\mathbf{P}}({\bf r})> = (\epsilon -1) <\widehat{\mathbf{E}}({\bf r}) >$.

In the simulations the external field  ${\bf {\cal E}}$ is chosen parallel to
the $Oy$ axis
and the microscopic field $\widehat{{\bf E}}_d({\bf r})$ is calculated for an ensemble of
$N_c=65{\rm x}65= 4225$ points located at the grid points 
of a square lattice.  It is calculated even if the point ${\bf r}$ is inside the hard
core of the disk. Hence the possibility of a rare but large contribution to $\widehat{{\bf E}}_d^{\delta}({\bf r})$, which
is avoided by setting to zero the values 
  $r<\delta =0.1 \sigma$ in the tabulation of the gaussian functions in Eq.(\ref{eq4}). Clearly it amounts
to compute 
\begin{equation}
 \widehat{{\bf E}}_d^{\delta}({\bf r})  = 2 \pi \sum_{i=1}^N \mathbf{G}_0^{\delta} (\mathbf{r},\mathbf{r}_i) \cdot {\bm \mu}^*_i 
\end{equation}
where the truncated Green function $ \mathbf{G}_0^{\delta}$ is that of Eq.~\eqref{decompo_b}.
It follows then from  Eq.~\eqref{decompo_a} that
    \begin{eqnarray}  
   < \widehat{{\bf E}}_d ({\bf r}) >  &=& <\widehat{{\bf E}}_d^{\delta}({\bf r})> -\pi <\widehat{{\bf P}} ({\bf r})> \, .
        \label{eq5}  
    \end{eqnarray}

 The Maxwell field has been computed  from Eq. (\ref{eq5}) for the thermodynamic states given in Table I.
For dipole moment  $\mu^*=1.5$, and all densities considered, the system
reaches  equilibrium after of the order 
of $10^5$ MC trial moves per particle and about $10^6$ MC trial moves per
particle appear sufficient to evaluate the canonical averages with a precision of
 $1\%$. Such a sampling is also sufficient to achieve  average values 
of $\widehat{{\bf E}}_d^{\delta}({\bf r}_s)$ at the different grid points $1,\, 2, \, ...\, , s, \, ...\, , N_c$
independent of ${\bf r}_s$ for all considered values of the external field. 

For dipole moments $\mu^*=2.5$ or $2.75$, a configuration of the system
is typically characterized by the formation of chains and rings. Such an
arrangement  of the disks evolves  to a different but similar
arrangement only within $10^5$ trial moves per particle and therefore 
of the order of  $10^7$ trial moves per particle are necessary for the estimate   
of the canonical averages and to obtain an average value of $ \widehat{{\bf E}}_d^{\delta} ({\bf r}_s)$  
quasi-independent of  ${\bf r}_s$. Figures  1 provides snapshots of such an evolution
at ${\cal E}_y=0.1$ for $\mu^*=2.5$ and $\rho^*=0.3$.

 The results of Table I show that for
all thermodynamic states considered the average value $<\widehat{{\bf E}}^d ({\bf r}) >$  
 $\simeq 0 $ as $-\pi <\widehat{{\bf P}} ({\bf r})>$ cancels  $<\widehat{{\bf E}}_d^{\delta}({\bf r})>$
 in the  limit of statistical errors.
 This result allows to  obtain two estimates of  the dielectric constant  $\epsilon$ 
 in the limit of small external fields, one, denoted $\epsilon_{\cal E}$, by
    \begin{eqnarray}  
    \epsilon_{\cal E} -1= 2 \pi <\widehat{P}_y ({\bf r})>/{\cal E}_y 
    \label{eq6} 
    \end{eqnarray}
 the other, denoted  $\epsilon_P$, by
     \begin{eqnarray} 
     (\epsilon_P-1)/(\epsilon_P+1)=\pi <\widehat{P}_y ({\bf r})>/({\cal E}_y+<\widehat{E}_{d, y}^{\delta}({\bf r})>) 
      \label{eq7} 
    \end{eqnarray}

 \begin{table} 
 
\begin{center}
\begin{tabular}{| c | c | c |  c | c | c | c | c |}   
\hline  
$ {\cal E}_y$  &   $\rho^*$   &  $\mu^*$  &   $<\widehat{E}_{d, y} ^{\delta}({\bf r})>$ & 
 $ -\pi <\widehat{P}_y ({\bf r})>$  & $<\widehat{E}_y ({\bf r}) > $  & $\epsilon_{\cal E}$ & $\epsilon_P$    \\
\hline
 0.0000   &  0.10  &  2.75     &  0.400E-03    & -0.106E-04	 &  0.389E-03	 &            &             \\
 0.0000   &  0.10  &  2.75     &  0.552E+00    & -0.552E+00	 & -0.368E-03	 &            &             \\  
 0.1000   &  0.10  &  2.75     &  0.323E+00    & -0.323E+00	 &  0.253E-03	 &            &             \\ 
 0.1000   &  0.10  &  2.75     &  0.710E+00    & -0.711E+00	 & -0.691E-03	 &            &             \\  
\hline
 0.0050   &  0.30  &  2.50     &  0.906E-01    & -0.900E-01	 &  0.565E-03	 &  0.370E+02 &  0.334E+02  \\
 0.0075   &  0.30  &  2.50     &  0.162E+00    & -0.162E+00	 & -0.128E-04	 &  0.442E+02 &  0.442E+02  \\
 0.0100   &  0.30  &  2.50     &  0.178E+00    & -0.179E+00	 & -0.968E-03	 &  0.368E+02 &  0.406E+02  \\
 0.0125   &  0.30  &  2.50     &  0.220E+00    & -0.220E+00	 & -0.206E-03	 &  0.363E+02 &  0.369E+02  \\
 0.0150   &  0.30  &  2.50     &  0.268E+00    & -0.268E+00	 & -0.687E-03	 &  0.368E+02 &  0.385E+02  \\
 0.0175   &  0.30  &  2.50     &  0.330E+00    & -0.330E+00	 & -0.144E-03	 &  0.387E+02 &  0.391E+02  \\
\hline
 0.0250   &  0.10  &  1.50     &  0.124E-01    & -0.124E-01	 & -0.298E-04	 &  0.199E+01 &  0.199E+01  \\ 
 0.0500   &  0.10  &  1.50     &  0.248E-01    & -0.248E-01	 & -0.142E-04	 &  0.199E+01 &  0.199E+01  \\
 0.0750   &  0.10  &  1.50     &  0.373E-01    & -0.373E-01	 & -0.342E-05	 &  0.199E+01 &  0.199E+01  \\
 0.1000   &  0.10  &  1.50     &  0.494E-01    & -0.494E-01	 &  0.676E-05	 &  0.199E+01 &  0.199E+01  \\
 0.2000   &  0.10  &  1.50     &  0.962E-01    & -0.963E-01	 & -0.432E-04	 &  0.196E+01 &  0.196E+01  \\
\hline 
 \end{tabular}
\end{center}
\caption{Maxwell field and dielectric constant for thermodynamics states at low and high reduced dipole moments
and for densities $0.1$ and $0.3$. $\epsilon_{\cal E}$ and $\epsilon_P$ are computed from Eqs.  (\ref{eq6} ) and (\ref{eq7} ),
respectively. At $\mu^*=2.75$, the evaluation of $\epsilon$ is precluded by
the metastability of the thermodynamic states.}.
\end{table}
 
 For $\rho^*=0.1$ and $\mu^*=1.5$  the two estimates of $\epsilon$ are in excellent
 agreement for all values $ {\cal E}_y< 0.2 $. This value of $1.97$ is also in agreement with the value
$2.01$ determined by the fluctuation formula \cite{Caillol_Weis} at ${\cal E}=0$,
$\epsilon -1= \pi <\widehat{{\bf M}}^2>/A $ where $\widehat{{\bf M}}=\mu^*\sum_i \mathbf{s}_i$.
 At $\rho^*=0.3$ and $\mu^*=2.5$ the estimated value of $\epsilon$ is equal to $ \sim 38.0 \pm 2.0$ 
 if $0.01 <{\cal E}_y < 0.02$; for  $ {\cal E}_y < 0.01$
 the estimated values lie between $43.0$ and $33.0$, a dispersion corresponding
 to the fact that after more then $10^7$ trial moves 
 per particle the statistical error on  $<\widehat{P}_y ({\bf r})>$ remains of the
 order of $10\%$.  The  estimated value  is  in agreement with that, $38.0 \pm 1.0$, obtained
 from the fluctuation formula.
   
 At  $\rho^*=0.1$ and $\mu^*=2.75$, the field  $<\widehat{E}_{d,y} ({\bf r}) > $ is also
 zero in the limit of statistical errors, demonstrating 
that within the performed trial moves, the values of $<\widehat{P}_y({\bf r})>$ and 
$< \widehat{E}_{d,y}^{\delta}({\bf r}) > $
remain coherent such that the  Maxwell field remains equal to the external field.
However, it is manifest that for $\mu^*=2.75$, sampling of configurations is
affected by strong metastability as shown by comparison of the average
polarizations, obtained with ${{\cal E}_y}=0$ and $3\, 10^7$ trials per particle,
 using different initial conditions. In one of the runs, starting from an unpolarized
 initial configuration, the average final polarization is $\sim 10^{-5}$
 (first line of Table I) , whereas in a run
 starting from  an initial configuration with a polarization close to the maximal polarization
 $0.275$, the average final polarization is equal to $\sim 0.175$ (second
 line of Table I). Snapshots of characteristic configurations for these two states are given in Figure 2.
  Similarly at ${{\cal E}_y}=0.1$, for two
 runs with different initial configuration, the polarizations seem stable
 for a sampling of $3 \,  10^7$ trials per particle, with values $0.102$
and $0.226$, respectively. This important metastability affecting the polarization values precludes
a reliable  calculation of $\epsilon$ from Eqs. (\ref{eq6} ) or (\ref{eq7} ).

\section{Conclusion}
\label{Conclusion}

A 2D dipolar system with periodic boundary conditions in the presence
of a uniform external field behaves according to the laws of electrostatics.
This result allows to show by numerical simulation that the Maxwell field
inside the system is equal to the applied external field even if the system
is highly polarized and also for stable or possibly metastable states
where the configurations of the system are dominated by chaines and rings.
This equality between Maxwell and external field does, however, not permit
to overcome the difficulties inherent in the determination of the
dielectric constant of the system at low temperature. Indeed, at these
temperatures the simulations are affected by important metastability which
precludes a unambiguous determination of the stable thermodynamic state.

\newpage

\noindent


\begin{figure}[h!]
\centering
\begin{tabular}{ c c c }
\vspace{1.em} \\
\hspace{4.0em} & \includegraphics[scale=0.55]{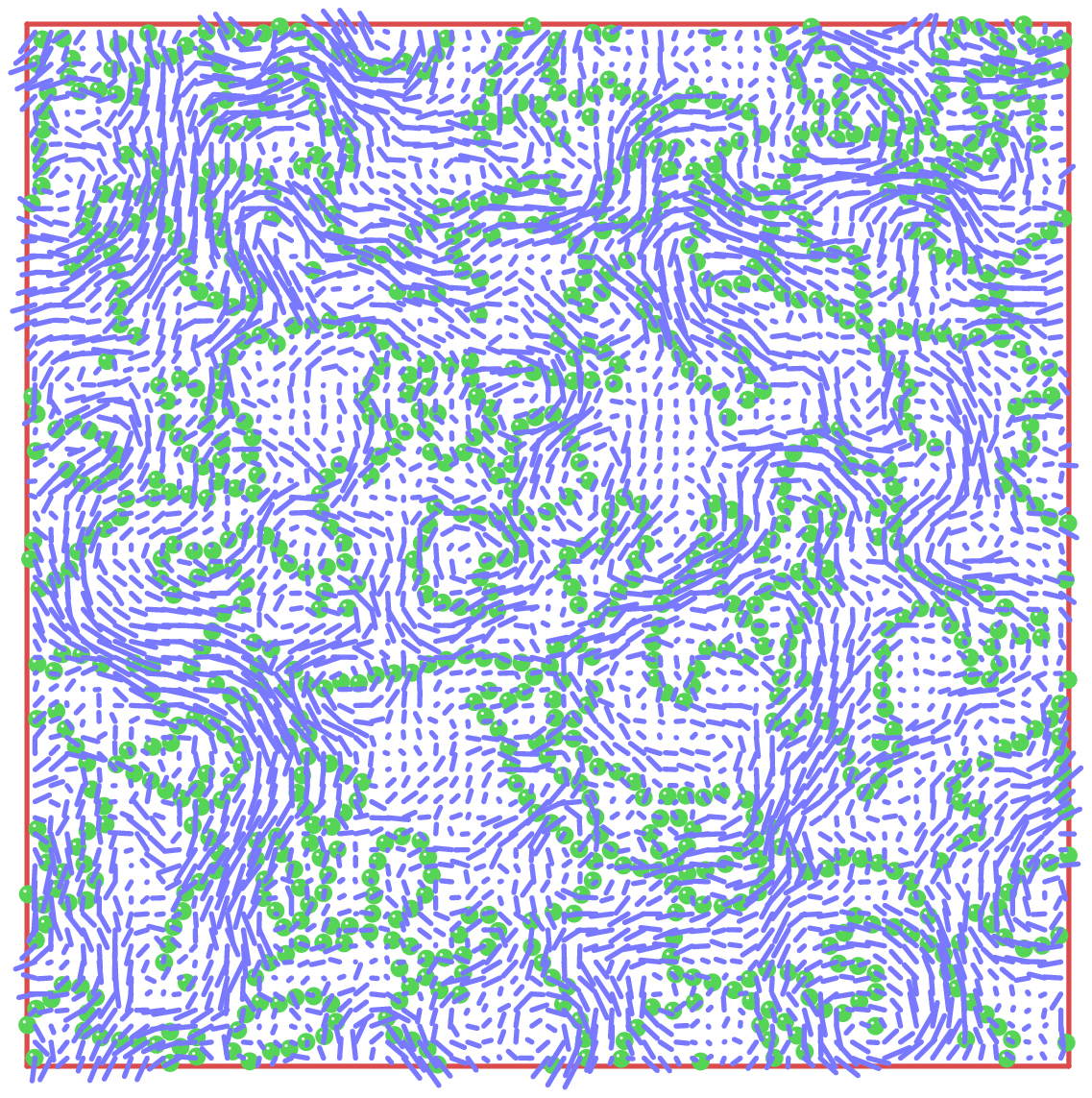} &   \includegraphics[scale=0.55]{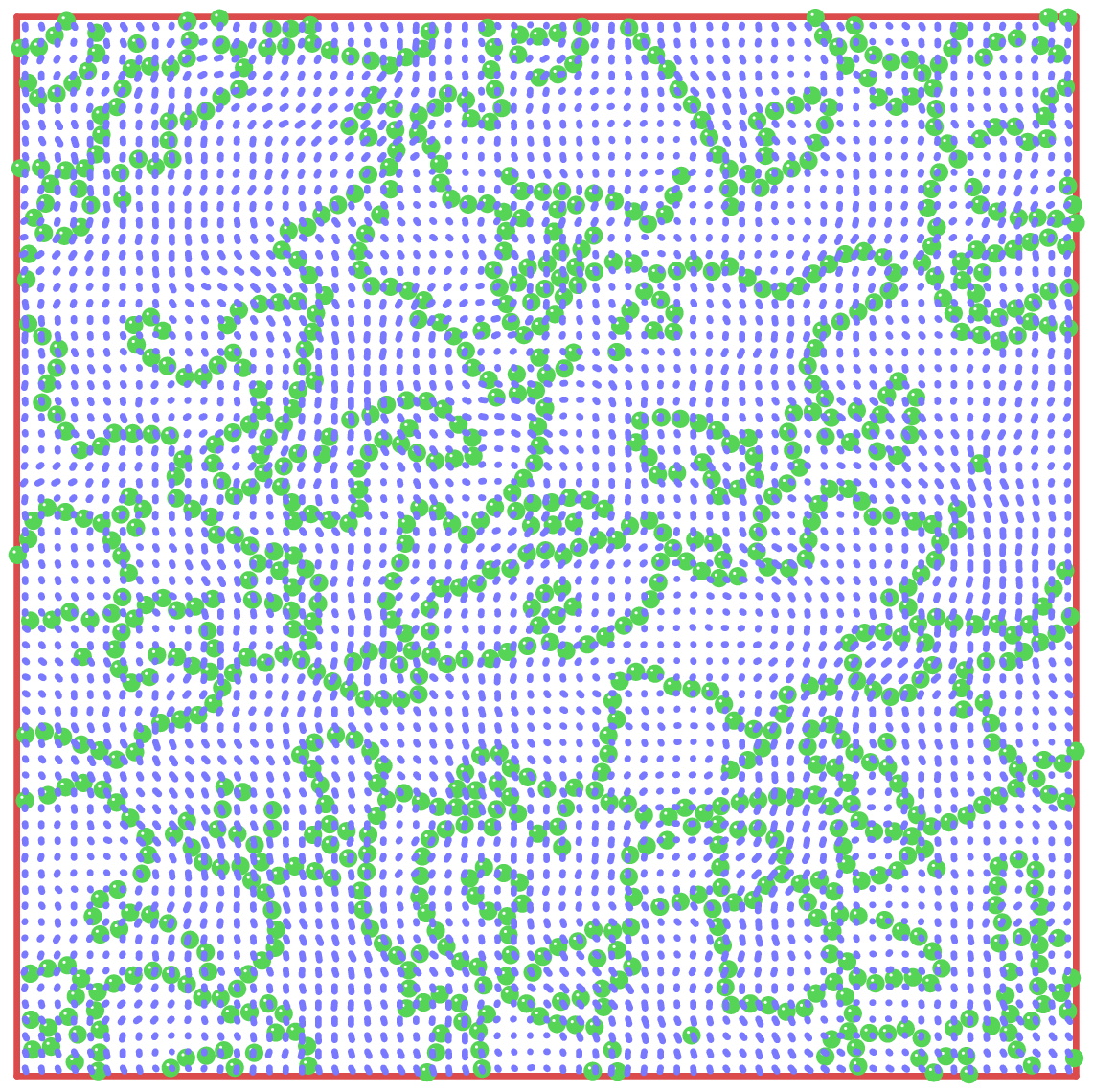}
\end{tabular}
\caption{Snapshots of  configurations of $1024$ dipolar discs at $\mu^*=2.5$, $\rho^*=0.3$ and 
 ${\cal E}_y=0.1$. The blue lines indicate the values and orientations of the average values of
  $ \widehat{{\bf E}}_d^{\delta} ({\bf r}_s)$
 at the different points ${\bf r}_s$.  Left : after $3\, 10^5$ trial moves per particle. 
  Right :  after $2\, 10^7$ trial moves per particle,
 the computed values $< \widehat{{\bf E}}_d^{\delta} ({\bf r}_s)>$ are quasi-uniform.}
\label{Fig1}
\end{figure}

\begin{figure}[h!]
\centering
\begin{tabular}{ c c c }
\vspace{1.em} \\
\hspace{4.0em} & \includegraphics[scale=0.55]{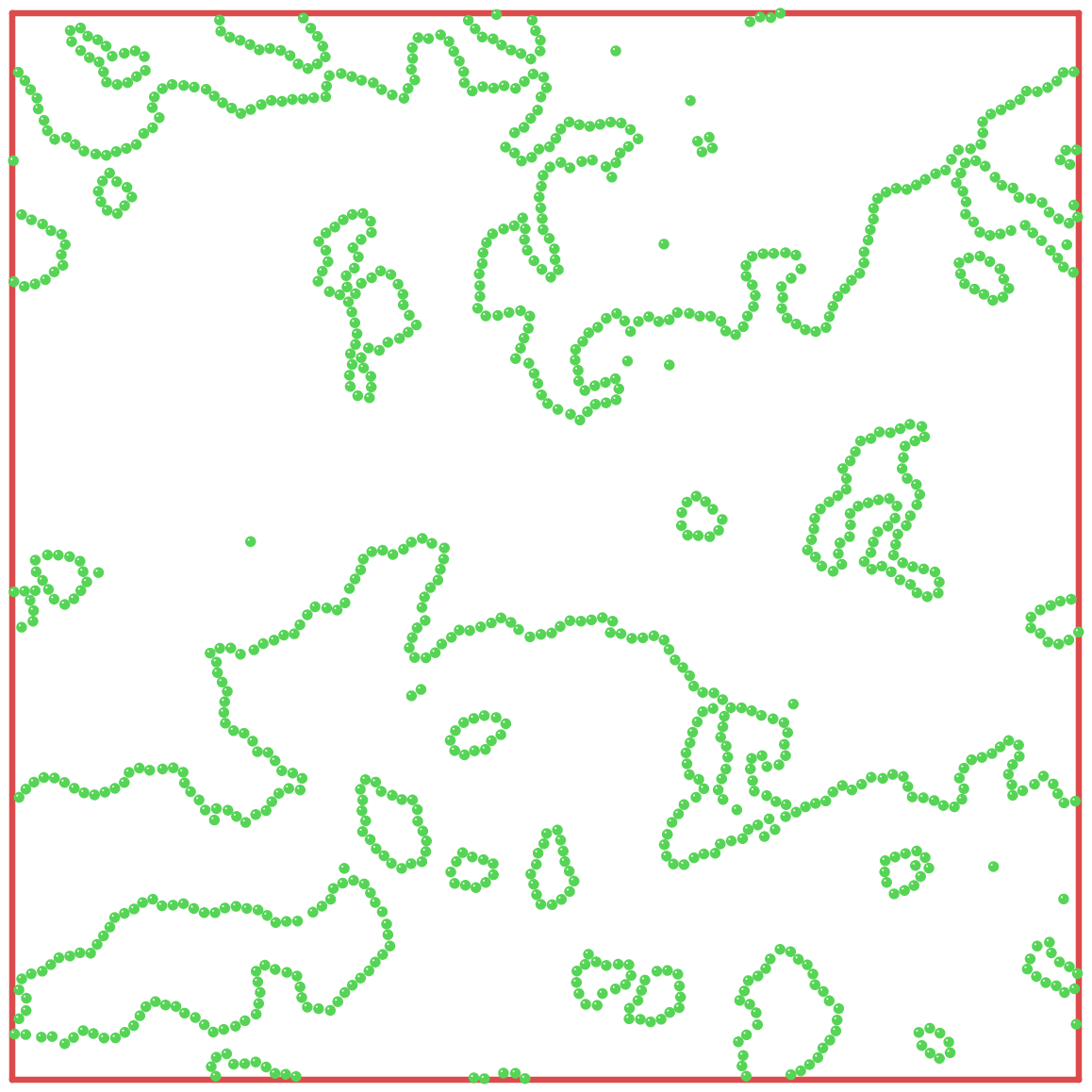} &  \includegraphics[scale=0.55]{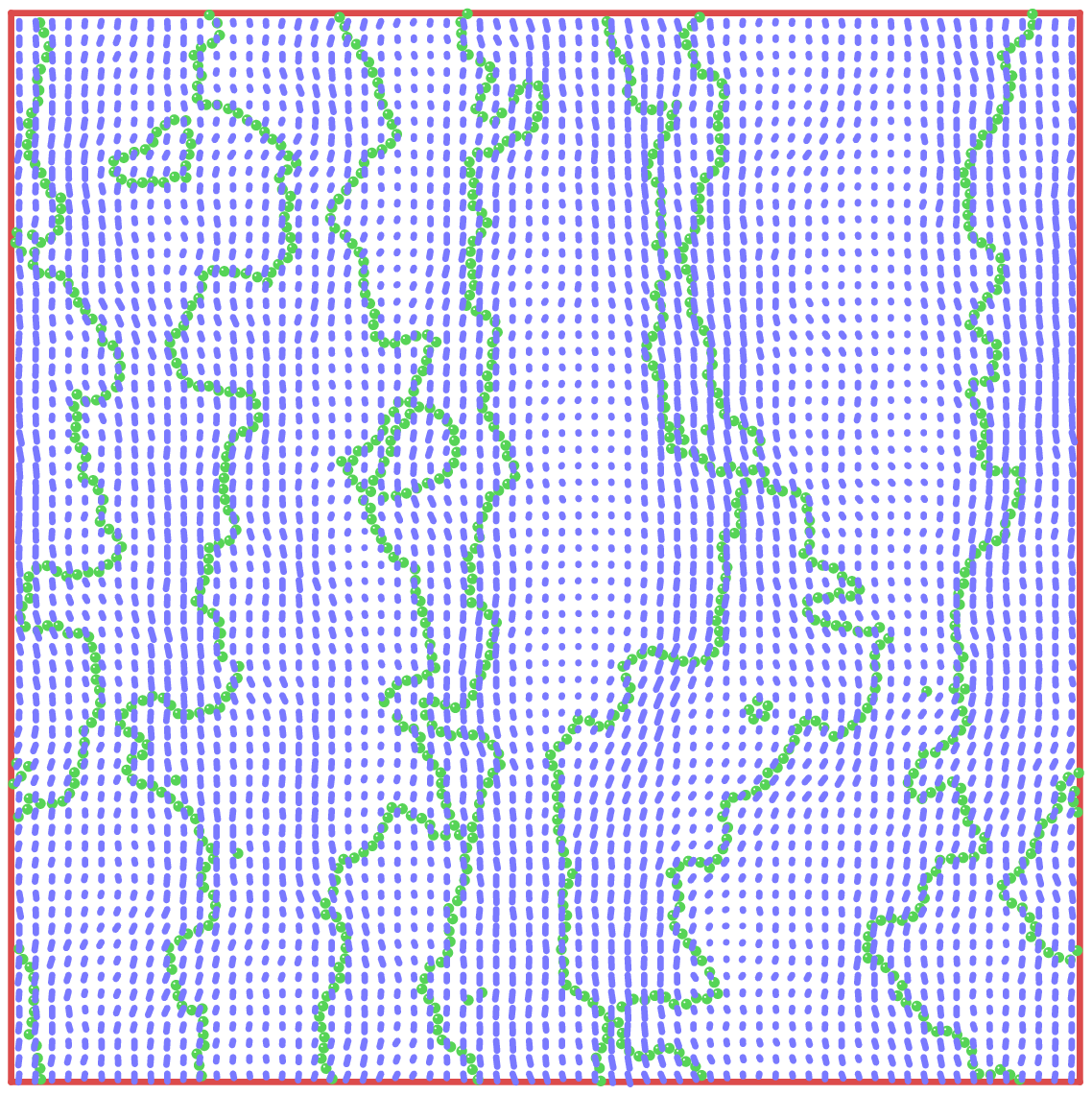}
\end{tabular}
\caption{Snapshots of  configurations of $1024$ dipolar discs at  $\mu^*=2.75$, $\rho^*=0.1$ and 
${\cal E}_y=0$ after $3\, 10^7$ moves per particle. Left : final configuration  for a run starting from
an initial unpolarized configuration; in this state $<\widehat{P}_y({\bf r)}>\simeq 0$ $< \widehat{{ E}}_{d,y}^{\delta} ({\bf r}_s)>\simeq 0$.
Right : final configuration  for a run starting from an initial polarized configuration;
 in this state the average polarization is $<\widehat{P}_y({\bf r)}>=0.175$ and $< \widehat{{ E}}_{d,y}^{\delta} ({\bf r}_s)>= -0.552$.}
\label{Fig2}
\end{figure}

\end{document}